\documentclass{ifacconf}

\usepackage{graphicx}      % include this line if your document contains figures
\usepackage{natbib}        % required for bibliography
\usepackage{subfigure}
\usepackage{amssymb, amsmath, amsfonts}
\usepackage{tikz}
\usepackage{pgfplots}
\pgfplotsset{compat=newest}
\usepackage{mathrsfs}
\usepackage{graphicx}
\usepackage{enumerate}
\usepackage[justification=centering]{caption}
\usepackage{url}
\usepackage{dsfont}

\usepackage{algorithm}
\usepackage{algpseudocode}
\graphicspath{ {./} }
\usetikzlibrary{matrix,arrows,fit,backgrounds,mindmap,plotmarks,decorations.pathreplacing}

\newlength\figureheight
\newlength\figurewidth
\newlength\fwidth

\makeatletter

\newcommand{\Rmnum}[1]{\expandafter\@slowromancap\romannumeral #1@}
\makeatother

\newcommand{\alert}[1]{\textcolor{black}{#1}}

\newcommand{\executeiffilenewer}[3]{%
	\ifnum\pdfstrcmp{\pdffilemoddate{#1}}%
	{\pdffilemoddate{#2}}>0%
	{\immediate\write18{#3}}\fi%
}

\newcommand{\qedliu}{\hfill{$\blacksquare$}}

\DeclareMathOperator*{\subject}{s.t.}%subject to
\DeclareMathOperator*{\tr}{tr}     % trace

\begin{document}
	\begin{frontmatter}
		
		\title{An Optimal Linear Attack Strategy on Remote State Estimation}
		
		\thanks[footnoteinfo]{This work is supported by the A*STAR Industrial Internet of Things Research Program, under the RIE2020 IAF-PP Grant A1788a0023, the Knut and Alice Wallenberg Foundation, the Swedish Foundation for Strategic	Research, and the Swedish Research Council. \textcopyright 2020. This work has been accepted to IFAC for publication under a Creative Commons Licence CC-BY-NC-ND.}

		\author[a,c]{Hanxiao Liu} 
		\author[b]{Yuqing Ni} 
		\author[a]{Lihua Xie}
		\author[c]{Karl Henrik Johansson}
		
		\address[a]{School of Electrical and Electronic Engineering, Nanyang Technological University, Singapore (e-mail: hanxiao001@ntu.edu.sg, elhxie@ntu.edu.sg).}
		\address[b]{Department of Electronic and Computer Engineering, Hong Kong University of Science and Technology, Clear Water Bay, Kowloon,	Hong Kong (email: yniac@connect.ust.hk)
		}
		\address[c]{School of Electrical Engineering and Computer Science, KTH Royal Institute of Technology, Sweden (email: kallej@kth.se)}
		
		\begin{abstract}                % Abstract of not more than 250 words.
			This work considers the problem of designing an attack strategy on remote state estimation under the condition of strict stealthiness and $\epsilon$-stealthiness of the attack. An attacker is assumed to be able to launch a linear attack to modify sensor data. A metric based on Kullback-Leibler divergence is adopted to quantify the stealthiness of the attack. We propose a generalized linear attack based on past attack signals and the latest innovation. We prove that the proposed approach can obtain an attack which can cause more estimation performance loss than linear attack strategies recently studied in the literature. The result thus provides a bound on the tradeoff between available information and attack performance, which is useful in the development of mitigation strategies. Finally, some numerical examples are given to evaluate the performance of the proposed strategy.
		\end{abstract}
		
		
	\end{frontmatter}
	%===============================================================================
	
	\section{Introduction}
	
	Cyber-Physical Systems (CPSs), which integrate computational elements and physical processes closely, are playing a more and more critical role in a large variety of fields which include transportation, power grid, military and environment. Most of them are of great importance to the normal operation of society and even to the whole nation. Any successful cyber-physical attacks will bring huge damages to critical infrastructure, human lives and properties, and even threaten the national security. Maroochy water breach in 2000~(\cite{slay2007lessons}), Stuxnet malware in 2010~(\cite{karnouskos2011stuxnet}), Ukraine power outage in 2015~(\cite{whitehead2017ukraine}) and other security incidents, motivate us to pay more attention to the security of CPSs.
	
	Recently, \alert{an enormous amount of research effort} has been devoted to designing detection algorithms and secure state estimation strategies to enhance the security of CPSs. \cite{mo2009secure} and \cite{mo2015physical} analyzed the effect of replay attacks, where the attackers do not know the system information and replay the recorded measurements\alert{,} and proposed a physical watermarking scheme to detect this kind of attacks. \cite{liu2014detecting} proposed the nuclear norm minimization approach and low rank matrix factorization approach to create a mechanism based on the properties of the nominal power grid to detect data injection attacks in a power grid. \cite{teixeira2012revealing} characterized the properties of zero dynamics attacks and provided necessary and sufficient conditions that the changes of inputs and outputs should satisfy to reveal attacks. \cite{fawzi2014secure} proposed a novel characterization of the maximum number of attacks that can be detected and provided an algorithm motivated by compressed sensing to estimate the state with attacks. 

%	From the standpoint of attackers, the problem regarding how to design an optimal attack strategy is also of great interest among researchers. \cite{mo2015performance} formulated a constrained control problem subject to the attacker's strategy and characterized the maximal perturbation. A linear quadratic function was employed to capture the attacker's control goal and constraints in~\cite{chen2017optimal}. The authors stated that the linear feedback is the optimal attack strategy under false alarm constraints and provided two algorithms to derive the optimal attack sequence. In~\cite{zhang2015optimal}, the problem regarding how to schedule Denial of Service (DoS) attack with \alert{energy-constrained} attackers was studied. The optimal attack schedule in a special scenario was proposed and the optimal attack schedule with both \alert{energy-constrained sensor and attacker} was also analyzed. 
	
	To the best of our knowledge, the concept of stealthiness of the attack was first introduced as 
	$\epsilon$-stealthiness based on KL divergence in~\cite{bai2015security,bai2017data}. The authors provided the corresponding $\epsilon$-stealthy attack strategy to induce the maximal performance degradation \alert{for a scalar system through data injection}. \cite{kung2016performance} generalized the above results to vector systems and pointed out the differences between scalar systems and vector systems. Furthermore, \cite{bai2017kalman} was devoted to seeking the optimal attack by compromising sensors' measurements. In this paper, we adopt the stealthiness metric employed in \cite{bai2015security,bai2017kalman}. Different from these works focusing on obtaining the maximal performance degradation and then deriving the corresponding attack strategy, we consider to maximize performance degradation given a specific linear attack type. Moreover, the performance degradation metric is slightly different from the above works.
	
	The linear integrity attack in our work was first proposed in~\cite{guo2016optimal}. An optimal linear attack policy was proposed to achieve the maximal performance degradation while not being detected. Some other extensions under different scenarios on this work could be found in \cite{8259281,guo2017consequence}. \cite{guo2018worst} also investigated this attack type in the detection framework based on KL divergence, which relaxed restrictions on false data detectors. However, since this type of attack only considers the latest information\alert{, it} may not be optimal from the viewpoint of attacker. Motivated by this point, we consider a more general attack type which combines the past attack information and the latest innovation. Moreover, we focus on the sequence detection instead of one-slot detection. 
	
	This work considers the problem of designing a general linear attack strategy on remote state estimation under the condition of different stealthiness from the standpoint of the attacker. Our work builds on the above works and focuses on designing a more general linear attack strategy. The main contributions of this paper are threefold:
	\begin{enumerate}
		\item We propose a more general linear attack type which employs the past attack data as well as the latest innovation and introduce the concept of $\epsilon$-stealthy attacks to characterize the stealthiness level of an attack.
		\item We present the optimal attack strategy to achieve the maximal performance degradation for two specific attacks with different stealthiness. 
		\item We prove that the proposed strategy performs better than the existing linear attack strategies in terms of performance degradation. Some numerical examples are provided to show this result.
	\end{enumerate}
	
	\emph{Notations:} $x_{k_1}^{k_2}$ is the sequence $\{x_{k_1}, x_{k_1+1}, \cdots, x_{k_2}\}$. \alert{The spectral radius} $\rho(A) = \max\{|\lambda_1|$, $|\lambda_2|, \cdots, |\lambda_n| \}$, where $\lambda_1, \cdots, \lambda_n$ are the eigenvalues of the matrix $A\in\mathbb{R}^{n\times n}$. $I_n$ denotes the identity matrix of order $n$.   
	
	\section{PROBLEM FORMULATION}\label{sec:problem}
\begin{figure}[h!]
		\centering
		\includegraphics[width=9cm]{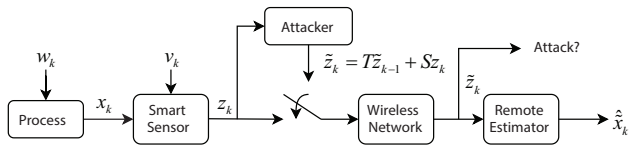}
		\caption{The system diagram.}
		\label{Fig:system}
	\end{figure}

	In this section, we introduce the system model as well as attack model. Besides, the stealthiness metric and performance degradation metric are provided to characterize the properties of attacks. Finally, we formulate the problem. The system diagram under consideration is illustrated in Fig. \ref{Fig:system}.
	
	\subsection{System Model}
	Let us consider a \alert{linear time-invariant (LTI)} system described by the following equations:
	\begin{align}
	x_{k+1} &= A x_{k} + w_{k}, \label{eq:systemdynamic}\\
	y_{k}  &= C x_k + v_k \label{eq:sensor}, 
	\end{align}	
	where $x_k\in \mathbb R^{n}$ and $y_{k}\in \mathbb{R}^{m}$ are the state vector and all sensors' measurement at time $k$, respectively. $w_{k}\in \mathbb{R}^{n}$ denotes the process noise and $v_{k}\in \mathbb{R}^{m}$ is the measurement noise. $w_k\sim \mathcal{N}(0,Q)$ and $v_k\sim \mathcal{N}(0,R)$, where $Q \geq 0$ and $R > 0$, respectively. It is assumed that $w_0,w_1,\ldots$ and $v_0,v_1,\ldots$ are mutually independent.%{\color{blue} Do not start a sentence with numbers or variables. Try to revise the sentences in this paragraph.}
	
	\begin{assum}\label{as:detectable}
		\alert{The spectral radius} $\rho (A)<1$ and the pair $ (A, C) $ is detectable and $(A, \sqrt{Q})$ is stabilizable.
	\end{assum} 
	
	%\begin{rem}
	%	Generally speaking, the control input is known by the controller and could be separated from the state update and sensor measurement equation. Hence, we do not consider it for the sake of simplicity.
	%\end{rem}

	The system is equipped with local smart sensors whose functions include signal conditioning, signal processing, and decision-making; see \cite{lewis2004wireless}. Here, we assume that the smart sensor employs the Kalman filter to process measurement and transmit the innovation to the remote estimator as follows:
	\begin{align*}
		&\hat{x}_{k+1|k} = A\hat{x}_{k|k},P_{k+1|k} = AP_{k|k}A^T+Q,\\
		&K_k = P_{k|k-1}C^T(CP_{k|k-1}C^T+R)^{-1},\\
		&\hat{x}_{k|k} = \hat{x}_{k|k-1} + K_k(y_k-C\hat{x}_{k|k-1}),\\ &P_{k|k} = P_{k|k-1}-K_kCP_{k|k-1},
	\end{align*}
	with initialization $\hat{x}_{0|-1} =\bar{x}_0$.
	
	It is known that the Kalman gain will converge exponentially due to {Assumption~\ref{as:detectable}}. Hence, we consider a steady-state Kalman filter with gain $K$ and the priori minimum mean square error (MMSE) $P$ for the remaining of this paper where 
	\begin{align}
		&P = \lim_{k\rightarrow \infty}P_{k|k-1},\label{eq:defineP}\\
		&K = PC^T(CPC^T+R)^{-1}.
	\end{align}
	Hence, the Kalman filter can be rewritten as:
	\begin{align*}
		\hat{x}_{k+1|k} = A\hat{x}_{k|k}, \hspace{0.3cm} \hat{x}_{k|k} = \hat{x}_{k|k-1} + K z_k,
	\end{align*}
	where $z_k \triangleq y_k-C\hat{x}_{k|k-1} $ is the innovation of the Kalman filter at time $k$, which will be transmitted to \alert{the remote estimator} and $z_k\sim \mathcal N(0,\sigma_z^2)$, where $\sigma_z^2 = CPC^T+R$.			
	
	\begin{rem}
		In our problem formulation, we assume that the innovation is transmitted to the remote estimator via a wireless communication network. Note that $y_k = z_k + C\hat{x}_{k|k-1}$, \alert{which means} $z_k$ contains the same information as $y_k$. In the existing works such as~\cite{ribeiro2006soi}, \cite{guo2016optimal}, \cite{li2017detection}, and \cite{8259281}, the sensor also sends innovation $z_k$ to the remote estimator.
	\end{rem}
	\subsection{Attack Model}
	Next we introduce the attack model. The adversary is assumed to have the following capabilities:
	\begin{enumerate}	
		\item The attacker has access to all the real-time {innovations} from smart sensors.
		\item The attacker can modify the true innovation to arbitrary {value} in a specific form.
		\item The attacker has the knowledge of system matrix $A$.
	\end{enumerate} 
	
	\begin{rem}
		The third capability could be relaxed. If the system parameter $A$ is not known, the attacker can learn it by system identification.% \alert{The identification accuracy affects the stealthiness and performance of the attack. Simulations in Section~\ref{sec:simulation} also show the attack effect with respect to different system dynamic matrices.} 
	\end{rem}
	
	The attacker records the real-time innovations from smart sensors and modifies them to $\tilde z_k$, \alert{i.e.,}
	\begin{align}
		\tilde z_{k} =  T \tilde z_{k-1} +  S z_k, \label{eq:linearattack}% + \Upsilon_k  \tilde{z}_{k-1} and $\Upsilon_k \in \mathbb{R}^m$ are
	\end{align}
	where $T \in {\mathbb{R}}^{m\times m}$ and $S \in \mathbb{R}^{m\times m}$.
	
	The remote estimator receives $\tilde{z}_k$ and \alert{updates the state estimate as follows:}
	\begin{align*}
		\hat {\tilde {x}}_{k+1|k} = A\hat{\tilde x}_{k|k}, \hspace{0.25cm} \hat{\tilde x}_{k|k} = \hat{\tilde x}_{k|k-1} + K\tilde{z}_k.
	\end{align*}
	Here, we initialize $\hat {\tilde {x}}_{0|-1} = \hat {{x}}_{0|-1} $ and $\tilde{z}_k = 0 $ for $ k\leq 0 $.
	\subsection{Stealthiness Metric}
	From the perspective of attackers, they should be stealthy or do not want to be detected by the system detector, otherwise the system will \alert{design countermeasures against attacks}. In this work, we employ a metric based on KL divergence measure to quantify the stealthiness of attack, which was first proposed in~\cite{bai2015security}. 
	
	Here, we propose the attack detection problem as a sequential hypothesis testing. The controller uses the received innovation sequence to carry out the following binary hypothesis testing:
	
	\textbf{$ \mathcal H_0: $} The remote estimator receives $z_1^k$.
	
	\textbf{$ \mathcal H_1: $} The remote estimator receives $\tilde z_1^k$.
	
	In testing $\mathcal H_0$ versus $\mathcal H_1$, there are two types of errors that can be made: the first type is called ``false alarm'', which denotes that the estimator decides $\mathcal H_1$ given $\mathcal H_0$, and the second type is called ``miss detection'', which represents that the estimator decides $\mathcal H_0$ when $\mathcal H_1$ is correct. Here, we denote the probability of miss detection at time $k$ as $p_k^M$, and the probability of false alarm is $p_k^F$. Furthermore, the probability of correct detection is $p_k^D$, which denotes that the controller decides $\mathcal H_1$ given $\mathcal H_1$. It is easy to know that  $p_k^D+p_k^M = 1$. Two definitions about attack stealthiness level are provided as follows:
	
	\begin{defn}[Strictly stealthy attack~(Bai et al., 2017a)] The attack is strictly stealthy if $p_k^F \geq p_k^D$ at time $k\geq 0$ holds for any detector.	
	\end{defn} 
	
	\begin{defn}[$\epsilon$-stealthy attack~~(Bai et al., 2017a)]\label{def:epsilon} The\\ attacker is $\epsilon$-stealthy if 
		\begin{align}
		\alert{\limsup_{k\rightarrow \infty}} -\frac{1}{k}\log p_k^F \leq \epsilon
		\end{align}
		holds for any detector that satisfies $ 0 < p_k^M<\delta $ for all times $ k $, where $0<\delta<1$.		
	\end{defn}
	
	\begin{rem}
	 Definition~\ref{def:epsilon} is motivated by Chernoff-Stein Lemma (see \cite{cover2012elements}). This lemma shows that the best exponent in probability of error is given by the relative entropy. Please refer to~\cite{bai2017kalman} for more details.
	\end{rem}

	\subsection{Performance Degradation Metric}
	In this paper, we employ the ratio of the state estimation error covariance $\tilde{P}$ and $P$ to quantify the performance degradation introduced by the attacker, i.e.,
%	\begin{align}
		$ \eta = \dfrac{\tr \tilde{P}}{\tr P}, $
%	\end{align}
	where $P$ is defined in \eqref{eq:defineP} and $\tilde{P}$ is defined as follows:
	\begin{align}\label{eq:tilde P}
		\tilde{P} \triangleq \alert{\limsup_{k\rightarrow \infty}}~ \frac{1}{k}\sum_{n=1}^{k}\tilde{P}_n,
	\end{align}
	where $\tilde{P}_n = E[(x_n-\hat{\tilde x}_{n|n-1})(x_n-\hat{\tilde x}_{n|n-1})^T]$\footnote{Akin performance degradation metric could be found in~\cite{bai2015security}.}.
	
	From the perspective of attackers, they need to design an appropriate attack strategy to maximize the ratio $\eta$, i.e., 
	\begin{equation}
		\begin{split}
		\arg_{T,S} &\hspace{0.5cm}\alert{\limsup\limits_{k\rightarrow\infty}}~ \dfrac{\frac{1}{k}\sum_{n=1}^{k}\tr \tilde{P}_n}{\tr P}.
		\end{split}
	\end{equation}
	\begin{rem}
		It is worth noticing that when there is no attack, $\tilde{z}_k = z_k$. As the initialization condition $\hat {\tilde {x}}_{0|-1} = \hat {{x}}_{0|-1} $, one can derive that $ \hat{\tilde x}_{k|k-1} =  \hat{x}_{k|k-1}$. Hence, $\tilde{P} = P$ and $\eta = 1$. In other words, the performance will not be degraded without attacks.
	\end{rem}
	
	\subsection{Problems of Interest}
	For the system described by~\eqref{eq:systemdynamic} and~\eqref{eq:sensor} under attack type~\eqref{eq:linearattack}, we aim to tackle the following two optimization problems:
	\begin{enumerate}
		\item 
		\begin{equation}
		\begin{split}\label{eq:op0}
			\hspace{0.3cm}	\max_{T,S} &\hspace{0.3cm}\alert{\limsup\limits_{k\rightarrow\infty} }~ \dfrac{\frac{1}{k}\sum_{n=1}^{k}\alert{\tr} \tilde{P}_n}{\alert{\tr}P},\\
			\subject &\hspace{0.3cm} \text{The attack is strictly stealthy.}
		\end{split}
		\end{equation}
		\item 	
		\begin{equation}
		\begin{split}\label{eq:op1}
			\hspace{0.3cm}	\max_{T,S} &\hspace{0.3cm}\alert{\limsup\limits_{k\rightarrow\infty}}~ \dfrac{\frac{1}{k}\sum_{n=1}^{k}\alert{\tr}\tilde{P}_n}{\alert{\tr}P},\\
			\subject &\hspace{0.3cm} \text{The attack is $\epsilon$-stealthy.}
		\end{split}
		\end{equation}
	\end{enumerate}
	
	We need to find the optimal attack pair $ (T^*, S^*) $ to induce the largest performance degradation while \alert{guaranteeing} that the stealthiness level satisfies the corresponding requirement. 
	
	\section{PRELIMINARY RESULTS}\label{sec:preliminary}
	In order to quantify the stealthiness level of attacks, we need to employ the KL divergence~(\cite{kullback1951information}, \cite{cover2012elements}), which is defined as:
	\begin{defn}[KL divergence] Let $x_1^k$ and $y_1^k$ {be} two random sequences with joint probability density functions $f_{x_1^k}$ and $f_{y_1^k}$, respectively. The KL divergence between $x_1^k$ and $y_1^k$ equals
		\begin{align}
		D(x_1^k\|y_1^k) =   \int_{-\infty}^{+\infty} \log \frac{f_{x_1^k}(\xi_1^k)}{f_{y_1^k}(\xi_1^k)}{f_{x_1^k}(\xi_1^k)}d \xi_1^k.
		\end{align}
	\end{defn}
	
	One can see that $D(x_1^k\|y_1^k) \geq 0$, {and $D(x_1^k\|y_1^k) = 0$ if and only if $f_{x_1^k} = f_{y_1^k}$}. Generally speaking, KL divergence is asymmetric, i.e.,  $D(x_1^k\|y_1^k) \neq D(y_1^k\|x_1^k)$. 
	
	The necessary and sufficient conditions for strictly stealthy attacks and $\epsilon$-stealthy attacks are provided as follows\footnote{For more details about the proofs, please refer to~\cite{bai2017kalman}.}: 
	\begin{lem}[Condition for Strictly Stealthy attacks]
		(Bai \newline et al. (2017a))
		\label{lem:strictstealthycond}
		An attack sequence ${\tilde z}_1^{\infty}$ is strictly stealthy if and only if ${\tilde z}_1^\infty$ is a sequence of i.i.d. Gaussian random variables with zero mean and variance $Cov(z_k) = CPC^T+R$.
	\end{lem}
	\begin{lem}[Conditions for $\epsilon$-stealthy attacks]
		(\cite{bai2017kalman})
		%~\cite{bai2017kalman}
		If an attack ${\tilde z}_1^\infty$ is $\epsilon$-stealthy, then 
%		\begin{align}
$$ 	\alert{\limsup_{k\rightarrow \infty}}~ \frac{1}{k}D( \tilde {z}_1^k\|z_1^k)\leq \epsilon. $$
%		\end{align}
		Conversely, if an attack sequence $\tilde {z}_1^\infty$ is ergodic and satisfies
%		\begin{align}
	$ 	\lim_{k\rightarrow \infty} \frac{1}{k}D(\tilde {z}_1^k\|z_1^k)\leq \epsilon, $
%		\end{align}
		then the attack is  $\epsilon$-stealthy.
	\end{lem}

	\section{MAIN RESULTS}\label{sec:main resrults}
	In this section, we will design an optimal attack strategy under strictly stealthy attacks and $\epsilon$-stealthy attacks. For the sake of analysis, we focus on the scalar case, i.e., $ m = n = 1 $. The vector case will be a potential future extension. The detailed solutions are provided in the following sections. 
	\subsection{Strictly Stealthy Attack}
	The goal of this subsection is to design an optimal attack pair $ (T^*, S^*) $ of the optimization problem~\eqref{eq:op0}. 
	\begin{thm}
		For a strictly stealthy attack, the optimal attack pair of the optimization problem~\eqref{eq:op0} is $(T^*,S^*)=(0, -1)$ and the corresponding performance degradation ratio is $\eta = 1+ \frac{4A^2K^2(C^2P+R)}{(1-A^2)P}$.
	\end{thm}
	
	\begin{pf}
		%Due to the space limit, we only give an outline of the proof. 
		Firstly, according to the attack type~\eqref{eq:linearattack} and the condition of strictly stealthy attacks in Lemma ~\ref{lem:strictstealthycond}, it is easy to derive that the feasible solutions of the problem~\eqref{eq:op0} are $(T,S)=(0, \pm 1)$.
				
		If $(T,S)=(0,1)$, $\tilde z_k=z_k$, which denotes that there is no attack in process, and the corresponding ratio $\eta =1$. 
		
		If $(T,S)=(0,-1)$, $\tilde z_k = -z_k$, which is aligned with the independence that strictly stealthy attack satisfies. Then we start to derive the corresponding ratio $\eta$. Rewrite $\tilde P_k$:
		\begin{equation}
		\begin{split}
			\tilde{P}_k =&E[(x_k-\hat{\tilde x}_{k|k-1})^2]\\
			%=&E[(x_k-\hat x_{k|k-1}+ \hat x_{k|k-1} -\hat{\tilde x}_{k|k-1})^2]\\
			=&P+E[(\hat x_{k|k-1} -\hat{\tilde x}_{k|k-1})^2]\\
			&+2 E[(x_k-\hat x_{k|k-1})(\hat x_{k|k-1} -\hat{\tilde x}_{k|k-1})]\\
			\overset{(a)}{=}&P+E[(\hat x_{k|k-1} -\hat{\tilde x}_{k|k-1})^2].
		\end{split}
		\end{equation}
		The reason why equation (a) holds can be found in~\cite{bai2017kalman}. We do not elaborate it here.
		
		Define $ \tilde{e}_k \triangleq \hat x_{k|k-1} -\hat{\tilde x}_{k|k-1} $. One can derive that 
		\begin{equation}
		\begin{split}
		\label{eq:b}
			E[\tilde{e}_k^2] &= E[(\hat x_{k|k-1} -\hat{\tilde x}_{k|k-1})^2], \\
			&\overset{(b)}{=}\sum_{n=1}^{k}A^{2n} K^2 E\left[\left[z_{k-n} -(-z_{k-n} )\right]^2\right]\\
			&=4A^2K^2(C^2P+R)\frac{1-A^{2k}}{1-A^2},
		\end{split}
		\end{equation}
		where (b) holds because $\tilde{e}_0=\hat x_{0|-1}-\hat{\tilde x}_{0|-1}=0$.

		Then, the covariance of the priori state estimate can be calculated as
		\begin{align*}
			\tilde{P} = \lim_{k\rightarrow \infty} \frac{1}{k}\sum_{n=1}^{k}\tilde{P}_n=P+ \frac{4A^2K^2(C^2P+R)}{1-A^2},
		\end{align*}
		and we obtain the performance degradation ratio:
		\begin{align*}
			\eta = \dfrac{P+ \frac{4A^2K^2(C^2P+R)}{1-A^2}}{P}= 1+ \frac{4A^2K^2(C^2P+R)}{(1-A^2)P}>1.
		\end{align*}
		Hence, the optimal attack pair of the optimization problem~\eqref{eq:op0} is $(T,S)=(0,-1)$ and the corresponding performance degradation ratio is $\eta = 1+ \frac{4A^2K^2(C^2P+R)}{(1-A^2)P}$. \qedliu
		
	\end{pf}
	
	\begin{rem}
		The above attack strategy is {aligned} with the results about the worst-case linear attack under the $\chi^2$ false alarm detector obtained in~\cite{guo2016optimal} and \cite{bai2017kalman}, which alter the sign of the innovation.
	\end{rem}

	\begin{rem}
		It is worth noticing that from \eqref{eq:b}, one can derive that $E[\tilde{e}_k^2] = 0$ for any attack if $A = 0$, and then $\eta = 1$. Hence, we do not consider this case for the later discussion.		
	\end{rem}
	
	\subsection{$\epsilon$-stealthy Attack}
	
	The goal of this subsection is to design an optimal attack pair $ (T^*, S^*) $ to maximize the ratio $\eta$, i.e., to maximize $ \tilde{P}$, %which can} achieve maximal performance degradation 
	under $\epsilon$-stealthy attacks.
	
	\begin{lem}\label{lem:detSigma}
		For any $T$, the differential entropy of the sequence $\tilde{z}_1^k$ can be expressed {as} $\frac{k}{2} \log (2\pi e S^2\sigma_z^2) $. 
	\end{lem}

	\begin{lem}\label{lemma:Tless1}
		If the attack is $\epsilon$-stealthy, then $|T|<1$.
	\end{lem}

	From Lemma \ref{lemma:Tless1}, for an $\epsilon$-stealthy attack, we have 
	\begin{align}\label{eq:simpbound}
	\lim_{k\rightarrow \infty} \frac{1}{k}D(\tilde {z}_1^k\|z_1^k) = -\frac{1}{2} - \frac{1}{2} \log( S^2 )+\frac{ S^2 }{2(1-T^2)}\leq \epsilon, 
	\end{align}
	where the third term of the above equation can be derived from the summation of geometric series.
	
	Then, we consider the performance degradation for {an} $\epsilon$-stealthy attack. An equivalent optimization problem is given in the following theorem, the proof of which can be found in the appendix for the sake of legibility.

	\begin{thm}\label{thm:equal1}
		The optimization problem \eqref{eq:op1} is equivalent to the following problem:
		\begin{subequations}
		\begin{align}\label{eq:op2}
			\max_{T, S} &\hspace{0.15cm} J(T,S)=(1-S)^2+ \frac{T^2  S^2 }{ 1-T^2 } - \frac{2ATS(1-S-T^2)}{(1-T^2)(1-AT)}, \\
			\subject &\hspace{0.15cm}-\frac{1}{2} - \frac{1}{2} \log( S^2 )+\frac{ S^2 }{2(1-T^2)} \leq \epsilon,\label{eq:constraint}\\
			&\hspace{0.5cm} |T|<1.
		\end{align}
		\end{subequations}
	\end{thm}	

	Next we seek to obtain a solution, i.e., an optimal attack pair $(T^*, S^*)$, of the above optimization problem. For the simplicity of notations, we use $J$ to denote $J(T,S)$.	
	
	First, let us consider the constraint condition~\eqref{eq:constraint}. Fix $T= T_o$ and $\epsilon = \epsilon_o\, (\epsilon_o \geq 0)$, $S$ must satisfy:
	\begin{align}\label{eq:reconstraint}
		-\frac{1}{2} - \frac{1}{2} \log( S^2 )+\frac{ S^2 }{2(1-T_o^2)} \leq \epsilon_o.
	\end{align}
	Define  
	\begin{align}\label{eq:redefineconstraint}
		\mathscr C(S,T_0,\epsilon_0)\triangleq -\frac{1}{2}-\frac{1}{2}\log(S^2)+\dfrac{S^2}{2(1-T_0^2)}-\epsilon_0.
	\end{align}	 
	It can be derived that only when $T_o^2\leq 1-e^{-2\epsilon_o}$, the inequality~\eqref{eq:reconstraint} has feasible solutions. And the solution lies in the interval $[-S_{\max}, -S_{\min}]\bigcup[S_{\min},S_{\max}]$, where $S_{\max}$ and $S_{min}$ are the largest and smallest positive solution {to} the equation $-\frac{1}{2} - \frac{1}{2} \log( S^2 )+\frac{ S^2 }{2(1-T_o^2)} = \epsilon_o$, respectively.

	\begin{lem}\label{lem:equality}
		The optimal attack pair $(T^*, S^*)$ must satisfy $-\frac{1}{2} - \frac{1}{2} \log( S^2 )+\frac{S^2 }{2(1-T^2)} = \epsilon$, where $S^*$ is the corresponding smallest solution for $T=T^*$. 
	\end{lem}

  \begin{rem}
  	Consider the property of the objective function, the optimal solution for $S$ is obtained when $S\leq 0$. Therefore, we only consider that $S\leq 0$ for the remaining of this work.
 	\end{rem}
	\begin{lem}
		When $S$ is negative and the absolute value of $T$ {is} fixed, $J(T,S)\geq J(-T,S)$, where the sign of $T$ is the same as the sign of $A$. 
	\end{lem}

	\begin{rem}
		For the simplicity of analysis, we only consider $T\geq 0$ and $A>0$. Hence, $T$ is non-negative in the above equation. The case when $T< 0$ and $A<0$ is essentially the same. 
	\end{rem}

	\begin{lem}
		The optimization problem \eqref{eq:op2} is equivalent to the following problem:
		\begin{subequations}
		\begin{align}
			\max_{S, T} &\hspace{0.15cm} J(T,S) = (1-S)^2+ \frac{T^2  S^2 }{ 1-T^2 } - \frac{ 2ATS(1-S-T^2)}{(1-T^2)(1-AT)} \nonumber\\
			\subject &\hspace{0.15cm}-\frac{1}{2} - \frac{1}{2} \log( S^2 )+\frac{ S^2 }{2(1-T^2)} = \epsilon\label{eq:constraint1op3} \\
			&\hspace{0.15cm} |T|\leq\sqrt{1-e^{-2\epsilon}},\label{eq:constraint2op3} 
		\end{align}
	\end{subequations}
	\end{lem}

	\begin{pf}
		It can be proved from~\eqref{eq:redefineconstraint} and Lemma~\ref{lem:equality}.\qedliu
	\end{pf}

	Rewrite \eqref{eq:constraint1op3} as follows:
	\begin{equation}\label{eq:TS}
		T=f(S)\triangleq\sqrt{1-\frac{S^2}{2\epsilon+1+\log (S^2)}}.
	\end{equation}
	
	Insert \eqref{eq:TS} into \eqref{eq:op2}, one has:
	\begin{equation}
	\small
	\begin{split}\nonumber
		J_1(S)=&-(2\epsilon+\log (S^2))- \frac{2S}{ 1-A f(S)} + \frac{2(2\epsilon+1+\log (S^2))}{1-A f(S)}\\
		& (-S_{o\max}\leq S\leq -e^{-\epsilon}),
	\end{split}
	\end{equation}
	where the range of $S$ is derived from $0 \leq T^2 \leq 1-e^{-2\epsilon}$.  

	\begin{thm}\label{theorem:strategy}
		The solution to the above optimization problem is an optimal attack pair $ (T_{opt}, S_{opt}) $, where $S_{opt}$ satisfies $\left.\frac{\partial J_1}{\partial S}\right|_{S=S_{opt}} = 0$ and {$T_{opt} = \sqrt{1-\frac{{S_{opt}}^2}{2\epsilon+1+\log (S_{opt}^2)}}$}. And the corresponding performance degradation ratio is $\eta = \dfrac{J_{opt} A^2K^2\sigma_z^2}{(1-A^2)P}$, where $J_{opt}= J(T_{opt}, S_{opt})$.
	\end{thm}
	\begin{pf}
		The main idea of the proof is to verify that the signs of the derivative of \alert{$J$} with respect to \alert{$S$} along the two boundaries are different, thus the optimal solution must exist in the feasible domain. 
		\begin{equation}
		\begin{split}\label{eq:partial J1}
		\frac{\partial J_1}{\partial S}
		=&\alert{-2}\dfrac{A^2f^2(S)+ S-ASf(S)-1 }{ S(1-Af(S))^2} \\
		&-2 \dfrac{\left[ S^2 -S(2\epsilon+1+\log S^2) \right]Af'(S)}{ S(1-Af(S))^2}, 
		\end{split}
		\end{equation}
		where 
		$ f'(S)=-\dfrac{\frac{S(2\epsilon+1+\log (S^2))-S}{ (2\epsilon+1+\log( S^2))^2}}{\sqrt{1-\frac{S^2}{2\epsilon+1+\log (S^2)}}}. $
		One can prove that when $S\rightarrow -S_{omax}$, the derivative of $J_1$ is positive. And the derivative at $S=-e^{-\epsilon}$ is negative.
	
			Since the function $J_1$ and the derivative of $J_1$ with respect to $S$ are continuous, there must be at least one maximum point where its first derivative is zero. Hence, $\eta = \dfrac{J_{opt} A^2K^2\sigma_z^2}{(1-A^2)P}$, where $J_{opt}= J(T_{opt}, S_{opt})$. \qedliu
	\end{pf}

	\begin{cor}
		The proposed attack strategy induces a larger performance degradation than the existing linear attack strategy in~\cite{guo2018worst} under \alert{the} same $\epsilon$-stealthy attacks. 
	\end{cor}

\section{Simulation}
\label{sec:simulation}
	In this section, we provide some numerical examples to evaluate the performance of the proposed attack strategy. We consider \alert{an} LTI system and set $A= 0.4, C = 1, Q = 0.2$, and $R = 0.5$. It is easy to derive that $K =0.3102$, and $P = 0.2248$. Here, we run $100$ thousand simulations to average them.
	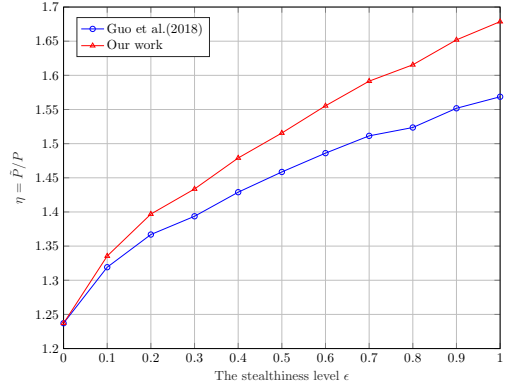
\begin{figure}[h!]		
		\begin{center}
					\begin{tikzpicture}[scale = 0.5]
					
					\begin{axis}[%
					width=4.521in,
					height=3.566in,
					at={(0.758in,0.481in)},
					scale only axis,
					xmin=0,
					xmax=1,
					xlabel style={font=\color{white!15!black}},
					xlabel={The stealthiness level $\epsilon$},
					ymin=1.2,
					ymax=1.7,
					ylabel style={font=\color{white!15!black}},
					ylabel={$\eta = \tilde P/ P$},
					axis background/.style={fill=white},
					xmajorgrids,
					ymajorgrids,
					legend style={at={(0.03,0.97)}, anchor=north west, legend cell align=left, align=left, draw=white!15!black}
					]
					\addplot [color=blue, mark=o, mark options={solid, blue}]
					table[row sep=crcr]{%
						0	1.2371\\
						0.1	1.319111644\\
						0.2	1.366930868\\
						0.3	1.393671731\\
						0.4	1.428926822\\
						0.5	1.458548168\\
						0.6	1.486098919\\
						0.7	1.511423874\\
						0.8	1.523626234\\
						0.9	1.551776956\\
						1	1.568642774\\
					};
					\addlegendentry{Guo et al.(2018)}
					
					\addplot [color=red, mark=triangle, mark options={solid, red}]
					table[row sep=crcr]{%
						0	1.2371\\
						0.1	1.335486801\\
						0.2	1.396889726\\
						0.3	1.433642534\\
						0.4	1.4790053\\
						0.5	1.51556642\\
						0.6	1.55538881\\
						0.7	1.591504522\\
						0.8	1.615452293\\
						0.9	1.651934091\\
						1	1.67862117\\
					};
					\addlegendentry{Our work}
					
					\end{axis}
					\end{tikzpicture}%
		\end{center}
				\caption{The ratio $\eta$ v.s. $\epsilon$ with fixed system parameters.}
	\label{fig:epsilonandeta}
	\end{figure}
	The ratio of the state estimation error covariance $\tilde{P}$ {to} $P$ v.s. stealthiness level $\epsilon$ is shown in Fig.~\ref{fig:epsilonandeta}. From this figure, one could see that the error covariance obtained in our work is equal to the one obtained in the exiting work \cite{guo2018worst} when $\epsilon = 0$. And the error covariance obtained in our work is larger than the one derived in \cite{guo2016optimal}. 
	Furthermore, the difference of the error covariance between our work and \cite{guo2018worst} \alert{gets} larger as $\epsilon$ grows.

	\section{Conclusion}\label{sec:conclusion}
	In this paper, an optimal linear attack strategy based on both the past attack signals and the latest innovation was proposed to achieve maximal performance degradation while guaranteeing a prescribed stealthiness level.  For strictly stealthy attacks, the result derived in this paper is aligned with the existing work. For $\epsilon$-stealthy attacks, we derived an optimal linear attack strategy and proved that the performance degradation of the optimal attack pair computed by using our proposed approach is better than the existing one. Simulation results were presented to support the theoretical results. For future works, we would like to generalize the results to a vector system, as well as analyze the performance of the optimal attack strategy under other performance metrics. 
	\section*{Appendix}
	\emph{Proof of Theorem~\ref{thm:equal1}:}
	\alert{Rewrite} $\tilde{e}_{k+1}$:
	\begin{equation}
	\begin{split}\label{eq:reek+1}
	\tilde{e}_{k+1} 
	&= A\tilde{e}_k + AK(1-S)z_k-AKT\tilde{z}_{k-1}.
	\end{split}
	\end{equation}
	From~\eqref{eq:reek+1}, we have $E[\tilde{e}_k] = 0$ and $\tilde e_k$ is independent of $z_k$. Hence, the covariance of $\tilde{e}_k$ is:
	\begin{align}
	&E[	(\tilde{e}_{k+1})^2]\nonumber\\
	=& A^2 E[	(\tilde{e}_k)^2]  + \left[AK(1-S) \right]^2\sigma_z^2 +(AKT)^2E[(\tilde{z}_{k-1})^2]\nonumber\\
	&+2A^2K(1-S)E[\tilde e_k z_k]- 2 A^2KTE[\tilde{e}_k  \tilde{z}_{k-1}]\nonumber\\
	\overset{(a)}{=}& A^2 E[	(\tilde{e}_k)^2]  + \left[AK(1-S) \right]^2\sigma_z^2 +(AKT)^2E[(\tilde{z}_{k-1})^2]\nonumber\\
	&- 2 A^2KTE[\tilde{e}_k \tilde{z}_{k-1}], \label{eq:tildeekcov}
	\end{align}
	where 
		\begin{equation}
	\begin{split}\nonumber
	\tilde{e}_{k} %&= \hat x_{k|k-1} -\hat{\tilde x}_{k|k-1}\\
	%&= A(\hat x_{k-1|k-2}- \hat{\tilde x}_{k-1|k-2}) + A K(z_{k-1}-\tilde{z}_{k-1})\\
	&= A\tilde{e}_{k-1} + AK(1-S)z_{k-1}-AKT\tilde{z}_{k-2}\\
	&= A^k \tilde{e}_0 + AK\left[\sum_{i=0}^{k-1}A^i(1-S)z_{k-1-i} -\sum_{i=0}^{k-1}A^iT\tilde{z}_{k-2-i}\right],
	\end{split}
	\end{equation}

	and Equation $(a)$ holds due to the independence \alert{between} $\tilde e_k$ and $z_k$.
	From~\eqref{eq:tildeekcov}, one can obtain that:
		\begin{align*}
		\lim\limits_{k\rightarrow \infty}\frac{1}{k}E[(\tilde{e}_{1})^2] =\lim\limits_{k\rightarrow \infty}\frac{1}{k}\left[AK(1-S) \right]^2\sigma_z^2\overset{(b)}{=}0,
		\end{align*}
		and 
		\begin{align*}
		\small
		&\lim\limits_{k\rightarrow \infty}\frac{1}{k}E[(\tilde{e}_{k+1})^2]\\ 
		\overset{(c)}{=}&\lim\limits_{k\rightarrow \infty}\frac{1}{k}\Big(  A^2 E[	(\tilde{e}_k)^2] -2 A^2KTE[\tilde{e}_k \tilde{z}_{k-1}]\Big)\overset{(d)}{=}0,
		\end{align*}
		where (b), (c) and (d) hold since $|A|<1$, $|T|<1$, $K$ and $\sigma_z^2$ are constants, and $S$ is bounded due to the property of~\eqref{eq:op2} and~\eqref{eq:constraint}.%{\color{blue} Does the boundedness exist due to Equation (11)?}
	
	Consider the asymptotic behavior for \eqref{eq:tildeekcov} and take the limit, we have:

	\begin{align*}
	\small
		&\lim\limits_{k\rightarrow \infty}\frac{1-A^2}{k}\sum_{n=1}^{k}E[(\tilde{e}_{n+1})^2]\\
		=&\left[AK(1-S) \right]^2\sigma_z^2 + \frac{(AKT)^2  S^2 }{ 1-T^2 } \sigma_z^2\\
		&- 2A^3K^2TS \left[\frac{ 1-S}{1-AT} -\frac{T^2S  }{(1-T^2)(1-AT)} \right]\sigma_z^2,
	\end{align*}
	Since $\sigma_z^2> 0$, $A^2K^2> 0$ and $P>0$, one can simplify the above optimization as 
	\begin{equation*}
	\small
	\begin{split}
		\max_{S, T} &\hspace{0.5cm} (1-S)^2+ \frac{T^2  S^2 }{ 1-T^2 } - \frac{2ATS( 1-S-T^2)}{(1-T^2)(1-AT)}, \\
		\subject &\hspace{0.5cm}-\frac{1}{2} - \frac{1}{2} \log( S^2 )+\frac{ S^2 }{2(1-T^2)} \leq \epsilon,\\
		&\hspace{0.5cm} |T|<1,
	\end{split}
	\end{equation*}
	where constraint conditions are from~\eqref{eq:simpbound} and Lemma
	\newline~\ref{lemma:Tless1}.
	\bibliographystyle{ifacconf}
	\bibliography{reference}
\end{document}